\documentclass{sigchi}

\toappear{\scriptsize Permission to make digital or hard copies of part or all of this work for personal or classroom use is granted without fee provided that copies are not made or distributed for profit or commercial advantage and that copies bear this notice and the full citation on the first page. Copyrights for third-party components of this work must be honored. For all other uses, contact the owner/author(s). Copyright is held by the author/owner(s). \\
{\emph{UIST'17 Adjunct, October 22--25, 2017, Qu\'ebec City, QC, Canada.} } \\
ACM ISBN 978-1-4503-5419-6/17/10. \\
http://dx.doi.org/10.1145/3131785.3131831}

\usepackage{balance}  
\usepackage{graphics} 
\usepackage{times}    
\usepackage{url}      
\usepackage[normalem]{ulem}     

\makeatletter
\def\url@leostyle{%
  \@ifundefined{selectfont}{\def\UrlFont{\sf}}{\def\UrlFont{\small\bf\ttfamily}}}
\makeatother
\def\pprw{8.5in}
\def\pprh{11in}

\usepackage[pdftex]{hyperref}
\hypersetup{
pdftitle={SIGCHI Conference Proceedings Format},
pdfauthor={LaTeX},
pdfkeywords={SIGCHI, proceedings, archival format},
bookmarksnumbered,
pdfstartview={FitH},
colorlinks,
citecolor=black,
filecolor=black,
linkcolor=black,
urlcolor=black,
breaklinks=true,
}

\begin{document}

\title{Non-Linear Editor for Text-Based Screencast}

\numberofauthors{1}
\author{
    \alignauthor Jungkook Park$^{1}$\thanks{Both authors contributed equally to this work.} , Yeong Hoon Park$^{2*}$, Alice Oh$^{1}$\\
     \affaddr{$^{1}$ School of Computing, KAIST, Republic of Korea}\\
     \affaddr{$^{2}$ Department for Computer Science \& Engineering, University of Minnesota, USA}\\
     \email{pjknkda@kaist.ac.kr}, \email{park1799@umn.edu}, \email{alice.oh@kaist.edu}
}

\maketitle

\begin{abstract}
Screencasts, where computer screen is broadcast to a large audience on the web, are becoming popular as an online educational tool. Among various types of screencast content, popular are the contents that involve text editing, including computer programming. There are emerging platforms that support such text-based screencasts by recording every character insertion/deletion from the creator and reconstructing its playback on the viewer's screen. However, these platforms lack rich support for creating and editing the screencast itself, mainly due to the difficulty of manipulating recorded text changes; the changes are tightly coupled in sequence, thus modifying arbitrary part of the sequence is not trivial.
We present a non-linear editing tool for text-based screencasts. With the proposed \textit{selective history rewrite} process, our editor allows users to substitute an arbitrary part of a text-based screencast while preserving overall consistency of the rest of the text-based screencast.

\end{abstract}


\category{H.5.m.}{Information Interfaces and Presentation (e.g. HCI)}{Miscellaneous}

\section{Introduction}

    Instructional screencasts are increasingly widespread as an online educational tool for a variety of topics. The medium of most screencasts is a video where the creator records their screen using a screen recording software. However, since the video is mainly a graphical, view-only medium, it is difficult to provide even a basic interaction with the in-video content such as dragging and copying a text shown in the video. To overcome this problem, there are emerging online platforms that support text-based screencast for demonstrating terminal sessions \cite{asciinema} or programming tutorials \cite{scrimba}. Instead of recording the screen as a video, these platforms capture the insertions/deletions in a character level, cursor/selection changes, or other relevant event data from the content creator's text editing activities to construct a text editing history, then provides the screencast to the viewers by reconstructing the text editing history with the data collected. This allows viewers to interact with the text in any given moment in the screencast, which is not possible with videos. For example, when viewing a programming tutorial screencast, a viewer may pause the playback, edit and run the code to better understand the content.
    
    Despite the advantage of text-based screencast, many of the currently available systems are lack of rich support for creating such content compared to other media such as video. In particular, little effort has been devoted to supporting \textit{non-linear editing} of text-based screencast, which is a method to randomly access and selectively edit intermediate parts of a content without a need to sequentially view and edit the content, which is a natural way of editing in the production of a video or audio contents today. 
    
    Unfortunately, implementing non-linear editing system for text editing history is technically challenging. Unlike video where each frame's data is independent of each other, each revision of a text editing history is dependent on all of its prior changes. For that reason, rewriting a part of a text editing history involves adjusting the offsets (numeric values indicating in which position the text is edited) of all subsequent changes. Moreover, rewriting a certain consecutive range of a text editing history can introduce ambiguity in reconstructing later versions of the text. While several studies previously introduced non-linear editing systems for document change history \cite{edwards1997timewarp} or code change history \cite{ginosar2013authoring, bird2009promises, hayashi2012refactoring}, most are based on snapshots or line-based diff system, which cannot give users a fine-grained control over the text editing history. 
    
    We propose \textit{selective history rewrite} process that enables substituting an arbitrary part of a text-based screencast while preserving overall consistency of text editing history. Then we present a web-based non-linear editor for text-based screencasts, and describe our user interface design for selecting history range from text editing history.


\section{Selective History Rewriting Process}
    
    A core component of our non-linear editor is the \textit{selective history rewriting} process that enables substituting a history range -- a consecutive part of a text editing history -- with a new history while preserving overall consistency. The process consists of two successive steps, \textit{validation step} and \textit{substitution step}, given a history range to be substituted and new text changes that will be placed in the history range.
    
\begin{figure}[tb!]
        \centering
        \includegraphics[width=0.4\textwidth, trim=7.5cm 7cm 7.5cm 6cm, clip]{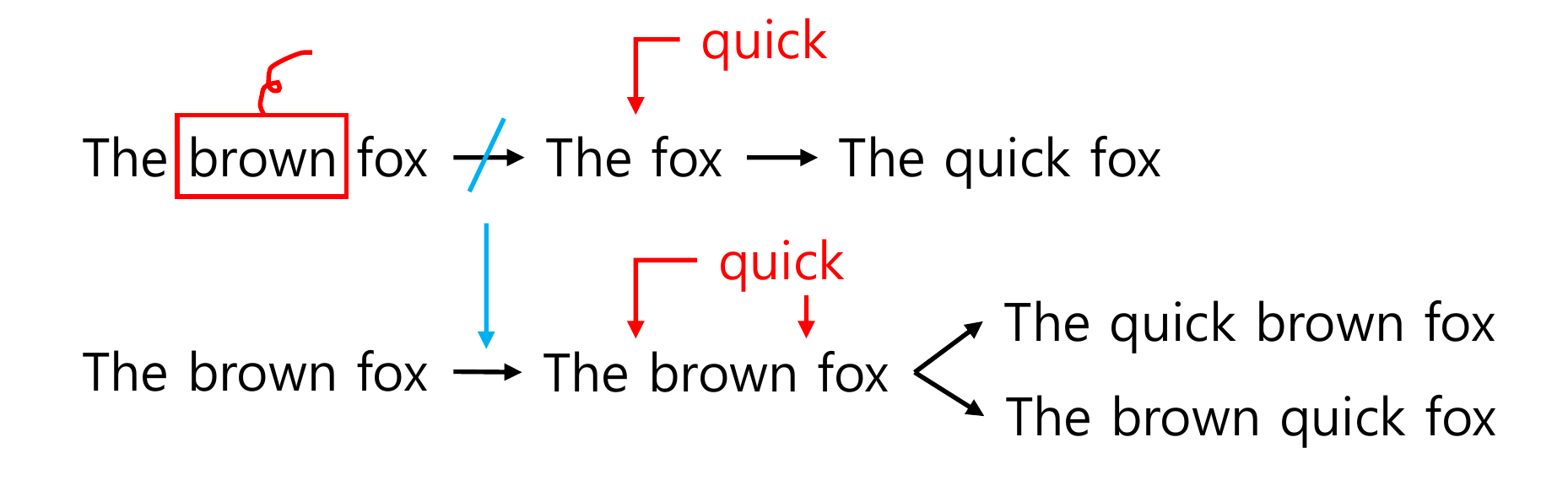}
    \caption{Removing the text change ``remove brown'' between the first and the second history introduces an ambiguity, resulting two valid sentences.}
        \label{fig:ambiguous_example}
    \end{figure}

    In \textit{validation step}, we evaluate whether rewriting a history range introduces any ambiguity on the subsequent part of the original history. To correctly validate the ambiguity, we calculate the ``effective area'' of the given history range and mark any text changes inside the effective area in the subsequent part of the history as ambiguous. When any ambiguous text changes is found, the rewriting process is aborted and user intervention is needed to change the selection. Figure~\ref{fig:ambiguous_example} shows one example which rewriting is not possible due to the violation in effective area.
    
    After the validation, we go through \textit{substitution step} where we calculate the effect of newly substituted history on the text editing history. Because the fore part of the original history are never affected by the substitution, we only re-calculate the offsets of the text changes in the subsequent part of the history. After the recalculation, we produce a new resulting screencast by combining the three partial text editing history -- (1) the fore part of the original history, (2) the newly substituted history, and (3) re-calculated history of the subsequent part.

    \begin{figure}[tb!]
        \centering
        \includegraphics[width=0.4\textwidth, trim=0 0 0 0.5cm, clip]{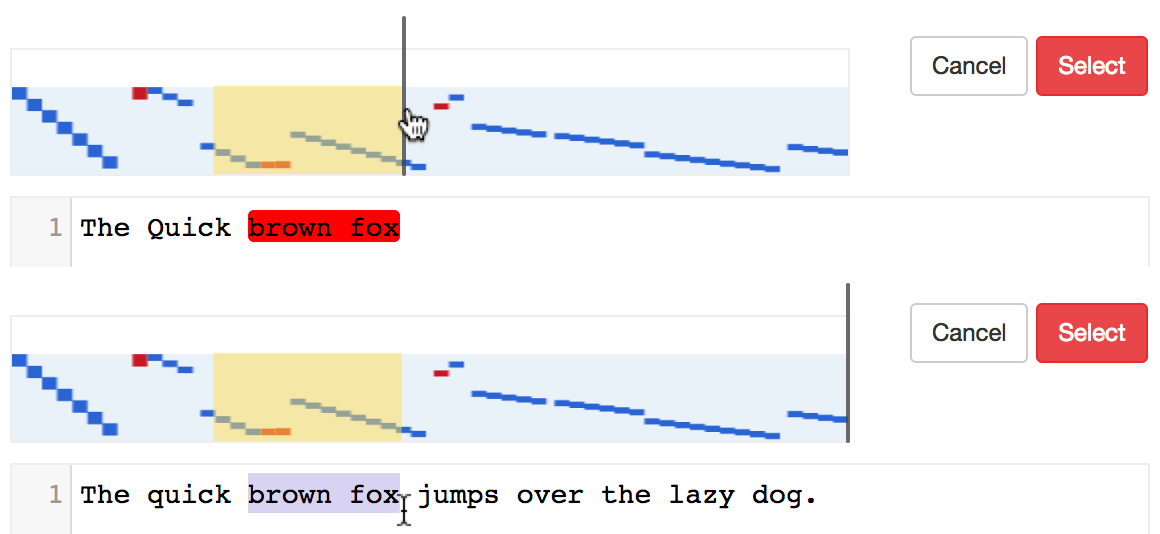}
    \caption{Selecting the same history range using two different selection method: Timeline-based selection (top) and Text-selection-based selection (bottom).}
        \label{fig:two_selection_methods}
    \end{figure}
        
    \begin{figure}[tb!]
        \centering
        \includegraphics[width=0.4\textwidth, trim=0 0 0 0.5cm, clip]{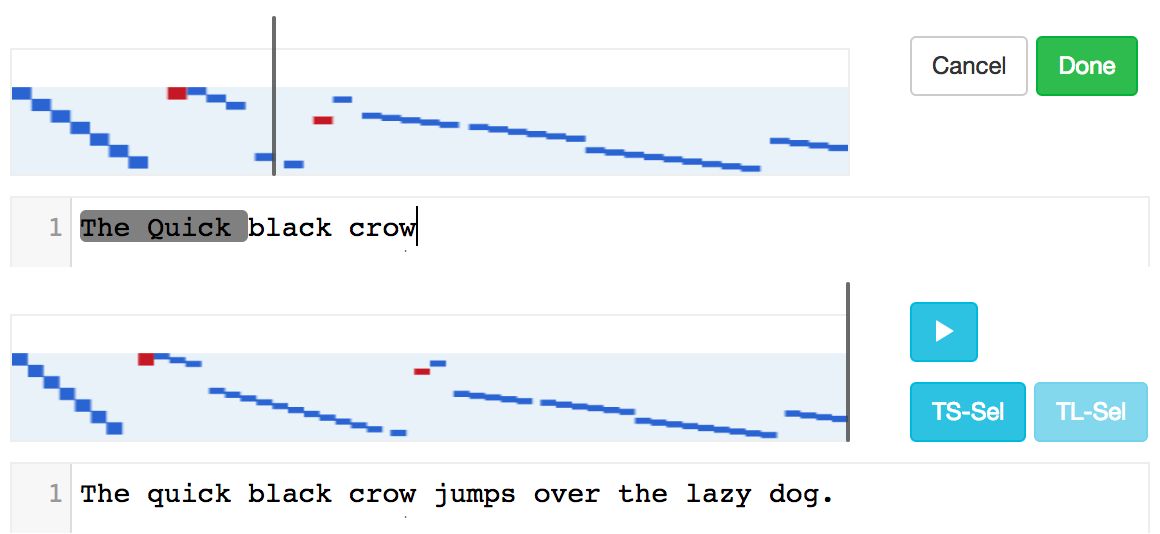}
        \caption{History rewrite mode (top) and the rewrited history with the string ``black crow'' in place of ``brown fox'' (bottom).}
        \label{fig:rewrite_history_editor}
    \end{figure}

\section{Text Editor with Selective History Rewriting}

    We built a web-based interface to demonstrate non-linear editing of text-based screencast and its user interaction. Our interface consists of two main components: \textit{History Slider} that visualizes the editing history (upper left in the interface), and the text area where user can view and edit the text. The \textit{History Slider} shows a timeline of the editing history along the x-axis, and the offset of text changes in y-axis.
    
    The foremost challenge in designing the interface for selective history rewriting was to find ways to allow users to easily select a history range they want to rewrite. We propose two interaction techniques for selecting a history range: \textit{Timeline-Based Selection (TL-Sel)} and \textit{Text-Selection-Based Selection (TS-Sel)}. Using either of the methods results in a selection of a contiguous history range. The selected range is visualized in the \textit{History Slider} as yellow highlighted region as shown in Figure~\ref{fig:two_selection_methods}.

    When user selects a history range using either \textit{TS-Sel} or \textit{TL-Sel}, the validation step of \textit{selective history rewriting process} immediately checks whether editing the selected history range introduces ambiguity. In case the selected history range is not rewritable, the editor disables the ``Select'' button to prevent user from starting rewriting the selected history. For \textit{TS-Sel}, the editor indicates when a selection is invalid by coloring the text selection in red.
    
    After selecting a valid history range, clicking the ``Select'' button enters the editor into edit mode. The text area now shows the earliest version of the text within the selected history range. In the text area, the user can only edit the regions that correspond to the effective area of the selected history range. In other words, the user can only edit the areas where the changes were made during the selected history range. The editor visualizes where the user can and cannot edit by showing gray background color behind the text that are not editable, as shown at the top of Figure~\ref{fig:rewrite_history_editor}. Clicking the ``Done'' button applies the substitution step of \textit{selective history rewriting process} and the editor returns to normal editing mode with the substituted history.

\section{Conclusion and Future Work}

    We introduced a web-based non-linear editor for text-based screencast, which allows users to substitute an arbitrary part of a text-based screencast while preserving overall consistency of the text editing history with the proposed \textit{selective history rewrite} process.
    
    
    
    In the current design, we simply prevent users from selecting a history range if rewriting the history range causes ambiguity. We believe it is crucial to devise with a way to resolve the ambiguity in order to make the editor practically usable for general use cases. One way would be to ask users to choose from the possible options. Another interesting idea is that some domain-specific heuristic algorithm or machine learning technique could be used to automatically identify which possible outcome is possibly the correct one so that we can eliminate the ambiguity.
    
    
    Moreover, more investigation is needed towards better visualization and navigation of text editing history. In our future study, we plan to find ways to improve the visualization to provide additional contextual information in the timeline, such as segmenting the timeline and labeling the segment with relevant information.

\balance


\bibliographystyle{acm-sigchi}
\bibliography{references}
\end{document}